\documentclass[cits]{PoS}

\usepackage{bm}
\usepackage{url}

\newcommand{\GeVc}{GeV${}/c$}

\newcommand{\Mgg}{M_{\gamma\gamma}}

\title{Forward Neutral Pion Cross Section and\\ Spin Asymmetry Measurements at STAR}
\ShortTitle{Forward Neutral Pions at STAR}

\abstract{The STAR endcap electromagnetic calorimeter (EEMC) was
  designed to allow measurement of cross sections and spin observables
  in the forward direction, $1 < \eta < 2$ and with the full
  azimuth. Using the EEMC to measure double longitudinal spin
  asymmetries in photonic channels---such as inclusive neutral pions,
  prompt photon, and prompt photon + jet---allows access to $\Delta G$
  within a low Bjorken-$x$ region ($0.01 < x < 0.33$ at $\sqrt{s}=200$) where $\Delta G$ is poorly
  constrained.  Additionally, transverse spin asymmetries, shown to be
  zero at $\eta$ near zero and as large as 10\% at $\eta$ near 4,
  occupy a previously unmeasured region in the 3D pseudorapidity,
  transverse momentum, and $x$-Feynman phase space when measured with
  the EEMC. The neutral pion cross section measurement verifies that
  the signal of interest can be properly reconstructed and isolated
  from the background. Pion reconstruction techniques in the STAR EEMC will be discussed and
  preliminary cross-section and transverse single spin asymmetry
  measurements presented.}

\FullConference{XXI International Workshop on Deep-Inelastic Scattering and Related Subjects\\
Marseille, France\\
22-26 April 2013}

\author{\speaker{S.~Gliske}$^{ab}$ and J.~Drachenberg$^{c}$, for the STAR Collaboration\\
\llap{$^a$}High Energy Physics Division\\
Argonne National Laboratory\\
Lemont, Illinois, USA\\
\llap{$^b$} University of Michigan\\
Ann Arbor, Michigan, USA\\
\llap{$^c$} Valparaiso University\\
Valparaiso, Indiana, USA\\
E-mail: \email{sgliske@umich.edu}, \email{jim.drachenberg@valpo.edu}}

\begin{document}

\section{Introduction}

The production of $\pi^0$-mesons in $\vec{p}+\vec{p}$ collisions at
$\sqrt{s}=200$ GeV accesses quark and gluon distributions
within the proton coupled with $\pi^0$ fragmentation functions.  At
intermediate pseudorapidity, the quark-gluon
sub-process dominates over gluon-gluon and quark-quark sub-processes, and
the lower Bjorken-$x$ is associated with the gluon.  The
measurements described in these proceedings, taken at an intermediate
pseudorapidity ($0.8 < \eta < 2.0$), cover previously unmeasured
regions of the $\eta$ and transverse momentum $p_T$ phase space.

The longitudinal double-spin asymmetry is sensitive to the gluon
polarization distribution $\Delta g(x)$. The longitudinal single-spin
asymmetries are parity violating and are thus expected to be very
small.  While $\Delta g(x)$ in the range $0.05 < x < 0.2$ is becoming
more constrained \cite{deltaGfit}, little is known for $x < 0.05$.
The analyzed data cover $0.01 < x < 0.33$ at $\sqrt{s}=200$. %, as shown in Fig. \ref{fig:x1x2}.
Preliminary results for the longitudinal
single-spin and double-spin asymmetries have been shown previously
\cite{Spin2010}.  This document will focus on the new cross-section
and transverse spin asymmetry results.

The measured cross section can be compared with perturbative QCD
(pQCD) calculations and add information regarding the gluon to $\pi^0$
fragmentation function.  Previous cross section measurements at nearby
kinematics \cite{STAR_pi0_BEMC, PHENIX_pi0_xSec} are typically within the pQCD prediction scale
uncertainty, lying at about 0.6 to 0.8 of the central scale prediction
in the region of $5 < p_T < 12$ \GeVc.

%\begin{figure}
%  \begin{center}
%    \includegraphics[width=0.49\textwidth]{bjorkenX.paper.eps}
%  \end{center}
%  \caption{\label{fig:x1x2}Distribution of Bjorken $x_1$ and $x_2$ for
%    $\pi^0s$ within EEMC acceptance for two different $\pi^0$ $p_T$
%    bins, based on a Pythia/GEANT simulation. }
%\end{figure}

Several processes may contribute to the transverse single-spin
asymmetry, $A_N$, for $x_F > 0$, including the Sivers and Collins
effects at twist-2 %\cite{AN_Siv_Col}
and higher twist effects. %\cite{AN_higher_twist}
Measurements in different kinematic
regions may help elucidate the underlying mechanisms.  The
$A_N$ measurements herein cover the previously
unmeasured region $0.06 < x_F < 0.27$ and $5 < p_T < 12$ \GeVc.

\section{Analysis}

The data were taken during the 2006 RHIC run using the STAR detector
\cite{STAR_NIM}.  A luminosity of 8.0 pb${}^{-1}$ was used for the
cross section results and 2.8 pb${}^{-1}$ for the transverse asymmetry
results.  The average transverse beam polarization was 0.55.  The
endcap electromagnetic calorimeter (EEMC) is used to measure the
energy and position of photons from $\pi^0$ decays.  The EEMC is a
lead-scintillator sampling calorimeter \cite{EEMC_NIM}, with 60
azimuthal and 12 radial segments (denoted towers).  Position
information is determined through a shower maximum detector (SMD),
consisting of two layers of tightly packed 1 cm wide scintillating
strips.  The response of each SMD plane is smoothed \cite{Tukey}, and
energy clusters are identified by a strip above 2 MeV with three
strips of monotonically decreasing energy on either side.  Clusters
from each of the two SMD planes are paired to determine the 2D
position on the EEMC, and energy deposited in the towers are used to
determine the energy of the incident particles, with a tower energy
resolution of $\delta E / E = 0.16 / \sqrt{E}$.  Pion candidates are
formed by making all possible pairs of incident particles passing
selection requirements.

\begin{figure}
\begin{center}
\includegraphics[width=0.49\textwidth]{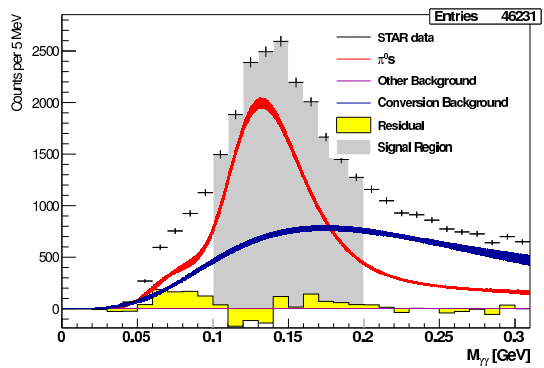}
\end{center}
\caption{\label{fig:mass}Invariant mass distribution for the
two-photon system with $8 < p_T < 9$ GeV${}/c$. Also included on the
plot are the template functions for the signal and two backgrounds
(scaled and shifted according to the fit results), the residual
between the data and the sum of the templates, and a gray-shaded area
indicating the signal region. }
\end{figure}

The $\pi^0$ signal fraction among the $\pi^0$ candidates was
determined by fitting a linear combination of template functions to
the two-photon invariant mass ($\Mgg$) distribution over the range
$0.0~<~\Mgg~<~0.3$ GeV$/{}c^2$ for each $p_T$ (or $x_F$) bin.  A typical bin is shown in Figure \ref{fig:mass}.  Three
template functions were determined by fitting Pythia
Monte Carlo data and represent (a) the $\pi^0$ signal, (b) the
conversion background where the two reconstructed ``photons'' of the
$\pi^0$ candidate were actually the two leptons from a photon that
converted in upstream material, and (c) all other backgrounds,
including combinatorial backgrounds.  When fitting the weights of the
three template functions an additional factor was also included to
account for the energy scale difference between the data and the Monte
Carlo.
%  This energy scale difference was not simply
%related to the calibration, but was also affected by assumptions about
%the signal fraction used in the simulation.

%The data and template functions for the $8 < p_T < 9$ GeV/$c$ bin are
%shown in Fig. \ref{fig:mass}.  While the fits to determine the signal
%fraction cover $0 < \Mgg < 0.3$, only $\pi^0$ candidates with $\Mgg$
%in the range $0.1 < \Mgg < 0.2$ GeV${}/c^2$ (defined as the peak
%region) were used for the remainder of the analysis.  The signal
%fraction in the peak region was computed from the weights, the data
%vs. simulation energy scale factor, and integrals of the template
%functions.  The product of the signal fraction in the peak region and
%the number of $\pi^0$ counts within this region then gives the number
%of background-subtracted $\pi^0$s for the given bin.

To compute the cross section, the number of background-subtracted
$\pi^0$s was corrected for $p_T$-bin smearing by applying the inverse
of a smearing matrix, obtained from a Pythia/GEANT Monte Carlo
simulation.  The final cross section was then computed using
Eq. \ref{eq:crossSection},
\begin{equation}\
  \label{eq:crossSection}
  E \frac{d\sigma}{d\bm p^3} = \frac{1}{\Delta \phi\ \Delta \eta\ \Delta p_T}
  \frac{1}{\langle p_T \rangle} \frac{1}{\textnormal{BR}} \frac{1}{\epsilon}\frac{N}{\mathcal L},
\end{equation}
where $N$ is the corrected number of $\pi^0$s, $\mathcal{L}$ is the
sampled luminosity (including dead-time corrections), $\epsilon$ is
the product of reconstruction and trigger efficiencies, BR is the
branching ratio $\pi^0\rightarrow \gamma\gamma$ \cite{PDG}, $\langle
p_T \rangle$ is the average $p_T$ for the particular $p_T$ bin,
$\Delta p_T$ is the width of the $p_T$ bin, and $\Delta \phi = 2\pi$
and $\Delta \eta = 1.2$ are the $\phi$ and $\eta$ phase space factors.
The trigger efficiency is below 10\% for $\pi^0$s with $5 < p_T < 6$
GeV, and plateaus above 40\% at $p_T \approx 9$~GeV/$c$.  The
reconstruction efficiency is around 30\% for $5<p_T<9$~GeV/$c$, and
decreases to around 20\% for $12 < p_T < 16$~GeV/$c$.

The transverse spin asymmetry was computed by binning with respect to
$\phi$, the angle between the azimuthal angles of the $\pi^0$ and the
spin polarization vector.  The raw cross ratio $\mathcal E(\phi)$ was
computed per $\phi$ bin, using Eq. \ref{eq:mathcalAN1},
\begin{eqnarray}
  \label{eq:mathcalAN1}
  \mathcal E(\phi) &=&
  \frac{\sqrt{N^{\uparrow}\left(\phi\right)N^{\downarrow}\left(\phi+\pi\right)}-\sqrt{N^{\downarrow}\left(\phi\right)N^{\uparrow}\left(\phi+\pi\right)}}{\sqrt{N^{\uparrow}\left(\phi\right)N^{\downarrow}\left(\phi+\pi\right)}+\sqrt{N^{\downarrow}\left(\phi\right)N^{\uparrow}\left(\phi+\pi\right)}}
\end{eqnarray}
where $N$ is the number of counts and $\uparrow$, $\downarrow$ denote the spin direction.  In a right handed system where $\hat{z}$ is defined by the beam momentum, spin up (down) points in the positive (negative) $\hat{y}$ direction."
The quantity $\mathcal E_N(\phi)$ was fit to the equation
$C+\varepsilon\sin\phi$.  The background was subtracted from $\varepsilon$ using the
signal fraction per bin and an estimate of the background asymmetry,
computed as the average of the asymmetry in the two sideband
$M_{\gamma\gamma}$-mass regions.  The background asymmetry was found
to be less than $1\sigma$ from zero, with $\sigma \approx 0.01$.  The
final result for $A_N$ was obtained by dividing by the luminosity
weighted polarization.

\section{Results}

\begin{figure}
\begin{center}
\includegraphics[width=0.49\textwidth]{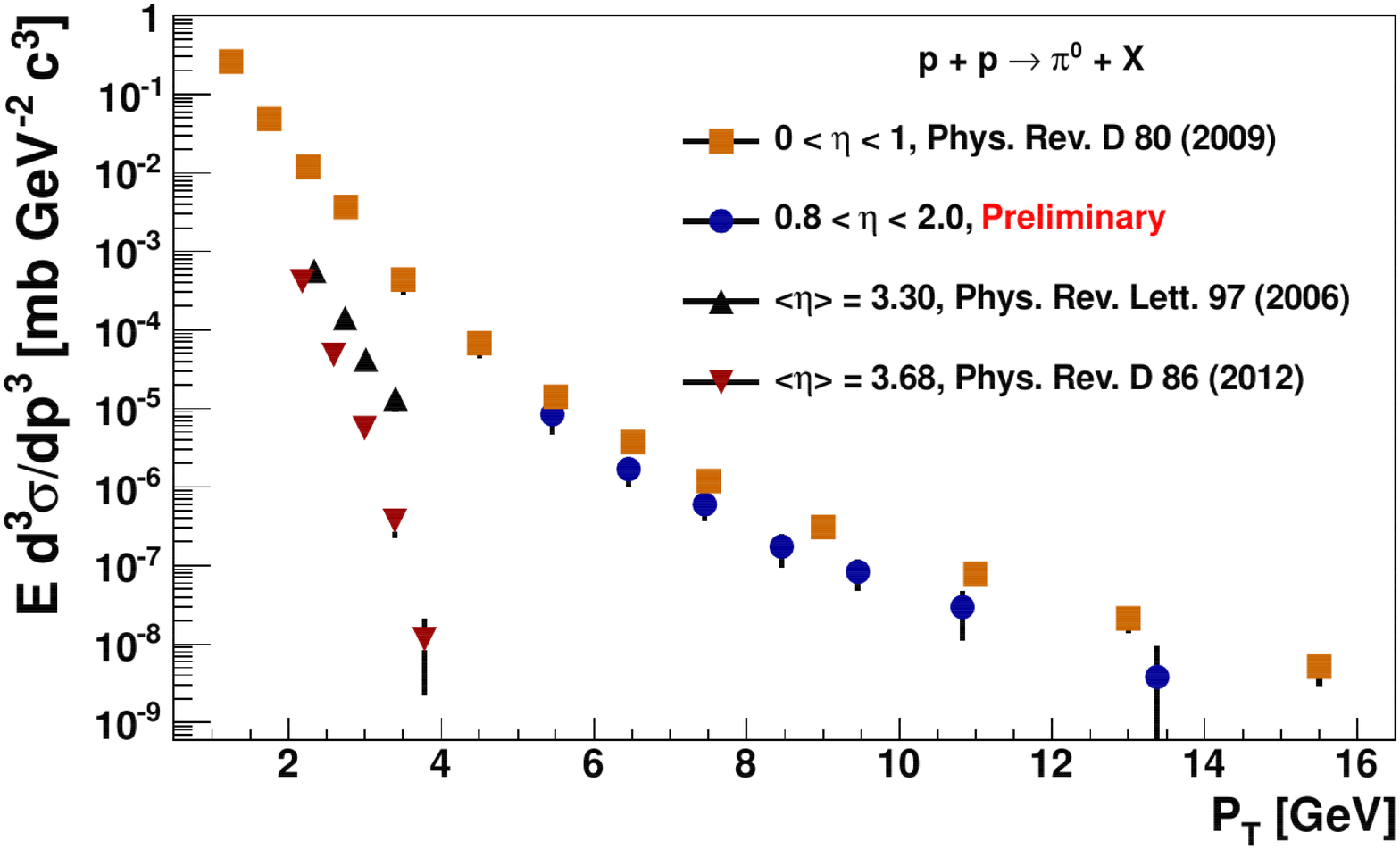}\ \ 
\includegraphics[width=0.49\textwidth]{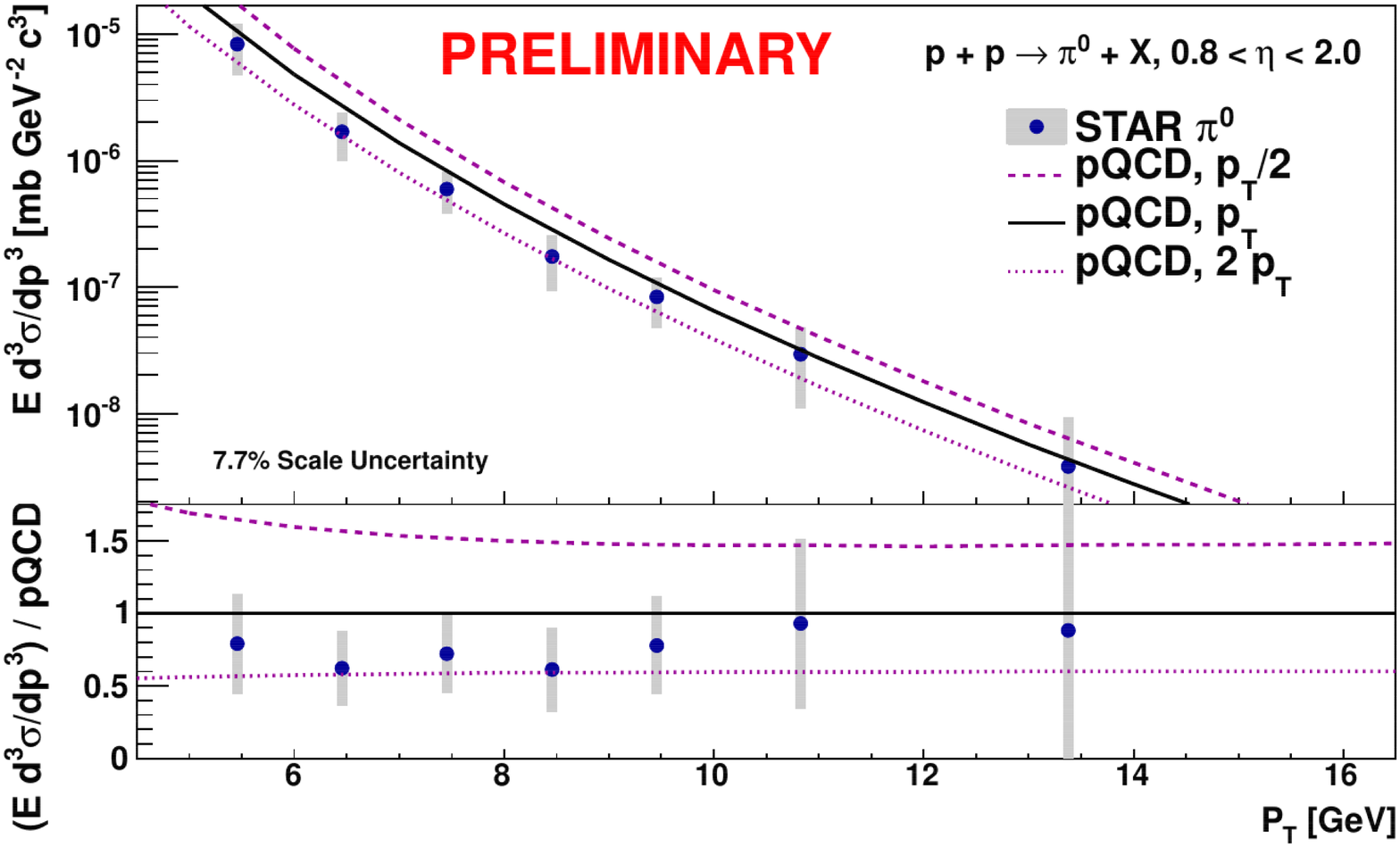}
\end{center}
\caption{\label{fig:xSec}Left Panel: $\pi^0$ cross section at various
  pseudorapidities measured by the STAR detector \cite{STAR_pi0_BEMC,STAR_FPD_xSec}.  Error bars are
  total uncertainty.
% The dark red squares are the results of this
%  analysis, while the other points are from Ref. \cite{STAR_pi0_BEMC}
%  (orange circles) and \cite{STAR_FPD_xSec} (blue and black
 % triangles).
  Right panel, upper section: $\pi^0$ cross section
  (black markers) is shown compared with a pQCD calculation
  \cite{xSec_Theory_Curve} with three options for the scale parameter.
  Statistical uncertainties are shown by the error bars, which are
  indistinguishable from the marker in most bins.  Systematic
  uncertainties are shown by the error boxes.  The right panel, lower
  section presents the ratio of the data to the theory curves for the
  various scales.}
\end{figure}

Figure \ref{fig:xSec} shows the cross section results of this analysis
in comparison with previously published STAR results in other
pseudorapidity and transverse-momentum regions, highlighting both the
broad spectrum of coverage of the STAR detector and that this result
lies in a previously unmeasured region.
%  Fig. \ref{fig:xSec} shows
%that the $\pi^0$ cross section is fairly flat with respect to $\eta$
%at lower $\eta$ and has significant $\eta$ dependence at higher
%$\eta$, with the transition lying between $\eta = 2$ and $\eta =
%3.68$.
In Fig. \ref{fig:xSec}, the cross section is also compared with a
theory curved based on pQCD and global fits of distribution and
fragmentation functions \cite{xSec_Theory_Curve}.  The EEMC $\pi^0$ cross section data points
are observed to lie between the $p_T$ and $2p_T$ scale.  This is
qualitatively consistent with published mid-rapidity STAR
\cite{STAR_pi0_BEMC} and PHENIX results at $\sqrt{s}=200$ GeV and
$\sqrt{s}=500$ GeV \cite{PHENIX_pi0_xSec} where
the cross section is lower than the $p_T$-scale theory curve in the
region of $6 < p_T < 16$ \GeVc. 
%Such a disagreement could indicate the
%importance of non-perturbative effects, or it may suggest the need for
%further refinements of the $\pi^0$ fragmentation function model.

%Contributions to the systematic uncertainties include the uncertainty
%on the signal fraction, the uncertainty on the smearing matrix, the
%effect of repeating the analysis with an additional $4 < p_T < 5$
%GeV/$c$ bin, the uncertainty on the reconstruction and trigger
%efficiencies, and the EEMC energy resolution and overall EEMC energy
%scale.  The signal fraction uncertainty includes contributions from
%the uncertainties of the parameters in the template functions, the
%uncertainty on the weights of the templates, the uncertainty on the
%scale parameter and its effect on the integrals used to determine the
%signal fraction in the peak, and a contribution based on the integral
%of the residual in the signal region.  The dominant uncertainty on the
%cross section is the overall energy scale uncertainty, which is
%correlated over all bins.

\begin{figure}
\begin{center}
\includegraphics[width=0.59\textwidth]{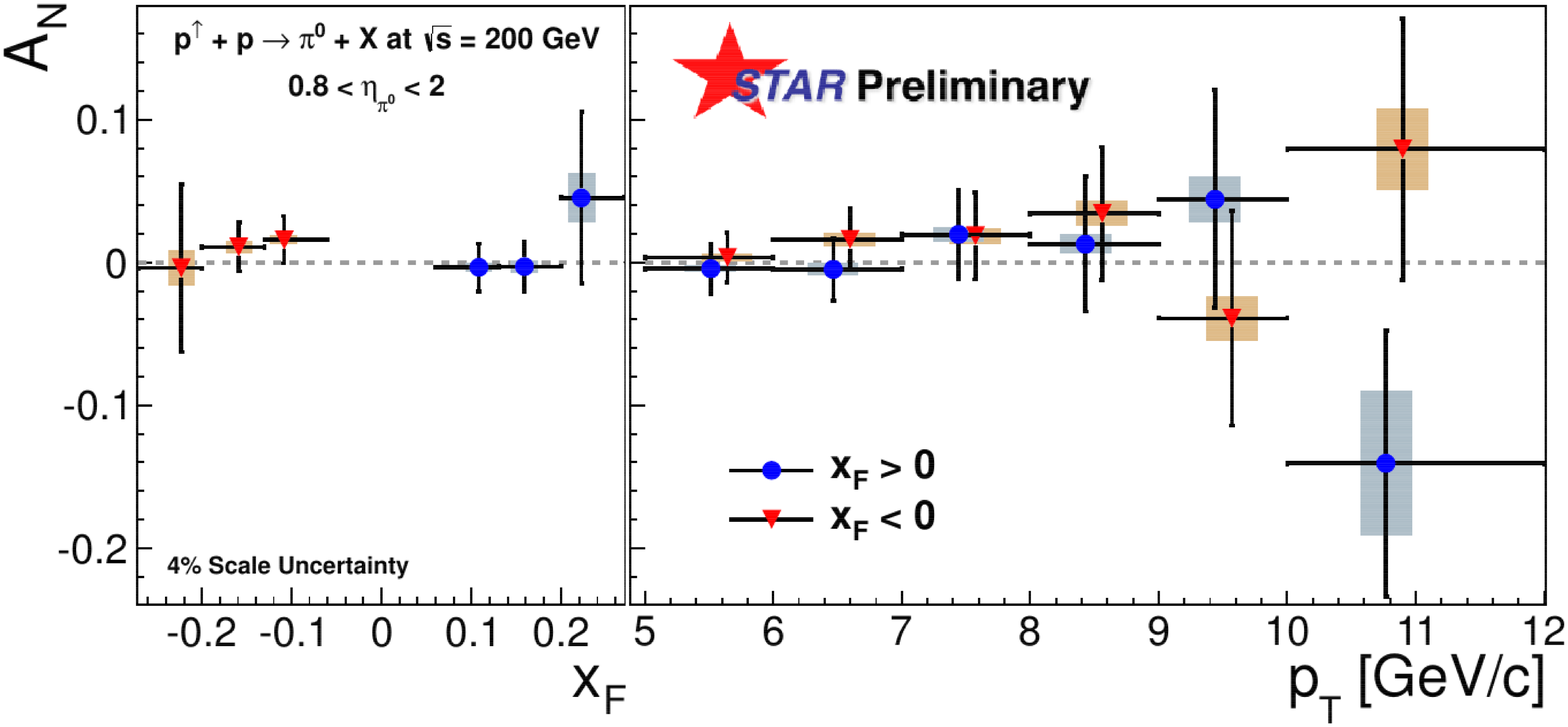}\ \ 
\includegraphics[width=0.39\textwidth]{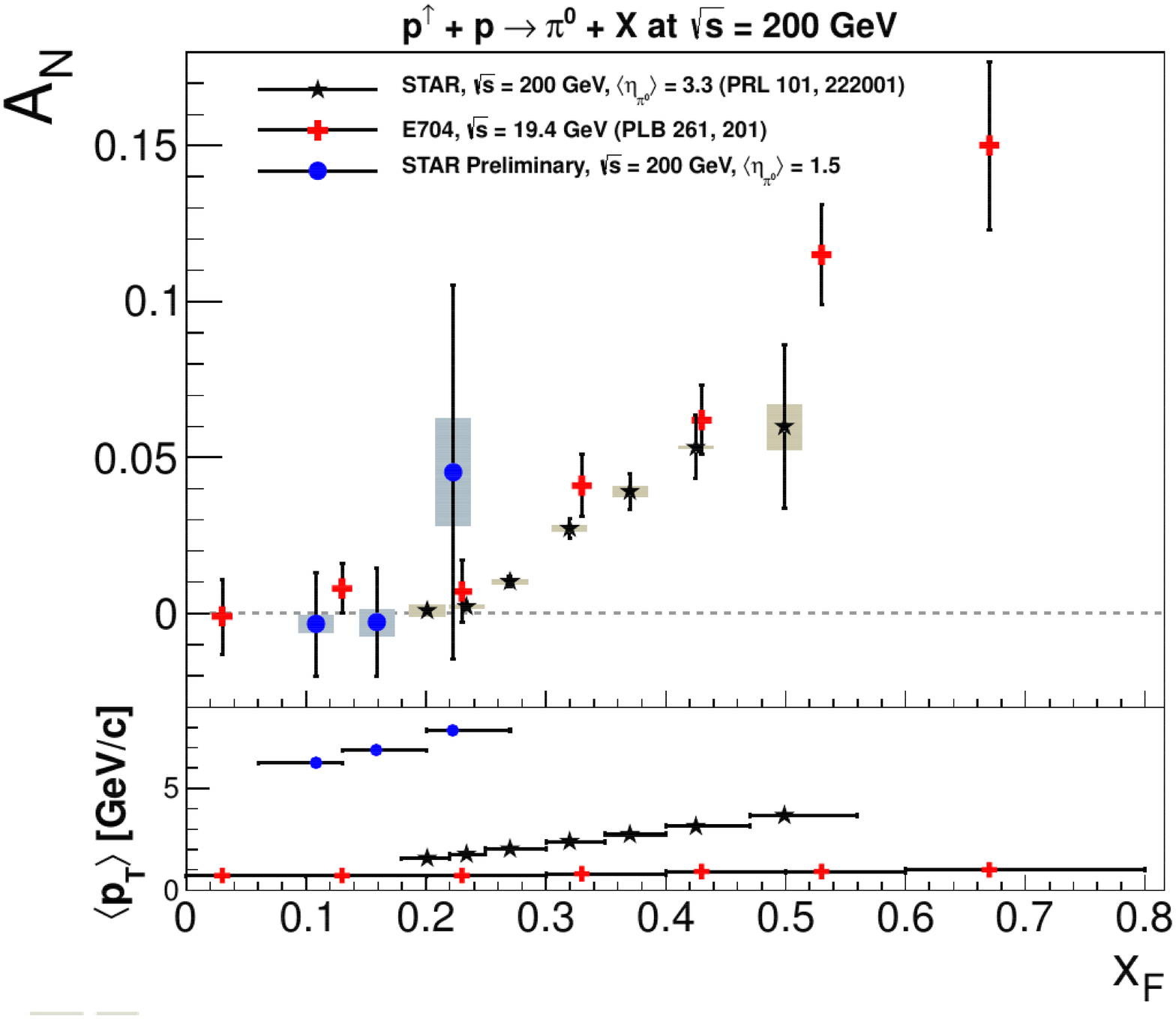}
\end{center}
\caption{\label{fig:AN} Left panel: $A_N$ results are plotted versus
  $x_F$ (left section) and versus $p_T$ (right section). Statistical
  uncertainties are shown by error bars, and systematic uncertainties
  are shown by error boxes. % Negative $x_F$ results are shown using
%  red triangular markers and tan systematic error boxes, while
%  positive $x_F$ results are shown using blue circle markers and blue
%  systematic error boxes.
  Right panel, top section: the $A_N$ results
  for $x_F>0$ versus $x_F$ are compared with previously published
  values of $A_N$.  Right panel, bottom section: the average $p_T$
  value is shown for each $x_F$ bin and for each experiment.
}
\end{figure}

The $A_N$ results versus $x_F$ for $0.06 < |x_F| < 0.27$ and $5 < p_T
< 12$ GeV/$c$ are shown in Fig. \ref{fig:AN}.  Systematic
uncertainties include the uncertainty on the signal fraction and
background asymmetry estimate, and single beam backgrounds.  $A_N$ is
consistent with zero for $x_F<0$, as expected.  As anticipated from
the previous results at lower $p_T$ and similar $x_F$
\cite{E704,STAR_FPD_AN}, $A_N$ is also consistent with zero for
$x_F>0$.  The $A_N$ results versus $p_T$, over the same range of $x_F$
and $p_T$, are also shown in Fig.  \ref{fig:AN}.  Within the $x_F$
region of this measurement, $A_N$ is
consistent with zero and no strong conclusions about the
$p_T$ dependence can be made.

%\section{Conclusions}

%Neutral pions were detected in the STAR Endcap Electromagnetic
%Calorimeter, having been produced in polarized proton-proton
%collisions with $\sqrt{s} = 200$ GeV at RHIC.  The production cross
%section, the double and single longitudinal spin asymmetries, and the
%single transverse spin asymmetry have been measured for $\pi^0$s with
%$0.8 < \eta < 2.0$ and with $5 < p_T < 12$ (spin asymmetries) or $5 <
%p_T < 16$ (cross section).  These results sample a region of phase
%space not previously studied, complementing measurements in
%neighboring regions of the phase space.  The cross section is slightly
%lower than previously published measurements at more central
%pseudorapidities and is within the scale uncertainty of a pQCD
%calculated prediction.  The measured values of the parity violating
%spin asymmetry, $A_N$ for $x_F < 0$, is consistent with zero.  The
%measured value of $A_N$ for $x_F > 0$ is also consistent with zero, as
%anticipated from previous results at lower $p_T$.  Final results for
%the cross section, the longitudinal spin and asymmetries and the
%transverse spin asymetries using these methods are in preperation.

\acknowledgments

The authors thank M. Stratmann for providing calculations and helpful
discussions.  We thank the RHIC Operations Group and RCF at BNL, our
national funding agencies, and the national funding agencies of our
collaborators.
% the
%NERSC Center at LBNL, the KISTI Center in Korea and the Open Science
%Grid consortium for providing resources and support. This work was
%supported in part by the Offices of NP and HEP within the U.S. DOE
%Office of Science, the U.S. NSF, CNRS/IN2P3, FAPESP CNPq of Brazil,
%Ministry of Ed. and Sci. of the Russian Federation, NNSFC, CAS, MoST
%and MoE of China, the Korean Research Foundation, GA and MSMT of the
%Czech Republic, FIAS of Germany, DAE, DST, and CSIR of India, National
%Science Centre of Poland, National Research Foundation
%(NRF-2012004024), Ministry of Sci., Ed. and Sports of the Rep. of
%Croatia, and RosAtom of Russia.
We also gratefully acknowledge a sponsored research grant for the 2006
run period from Renaissance Technologies Corporation.

\end{document}